\documentclass[aps,prb,twocolumn,footinbib]{revtex4}
\usepackage{amssymb,amsbsy,graphicx,times,subfigure,marvosym,color,bm,hyperref}
\usepackage[latin1]{inputenc}
\usepackage{hyperref}
\hypersetup{
    colorlinks,%
    citecolor=blue,%
    filecolor=black,%
    linkcolor=blue,%
    urlcolor=blue
}

\begin{document}

\title{Magnetic-coupling-dependent spin-triplet supercurrents in helimagnet/ferromagnet\\ Josephson junctions}
\author{G\'{a}bor B. Hal\'{a}sz}
\address{Department of Materials Science and Metallurgy, University of Cambridge, Pembroke Street, Cambridge CB2 3QZ, United Kingdom}
\author{M. G. Blamire}
\address{Department of Materials Science and Metallurgy, University of Cambridge, Pembroke Street, Cambridge CB2 3QZ, United Kingdom}
\author{J. W. A. Robinson}
\email{jjr33@cam.ac.uk}
\address{Department of Materials Science and Metallurgy, University of Cambridge, Pembroke Street, Cambridge CB2 3QZ, United Kingdom}


\begin{abstract}

The experimental achievements during the past year in demonstrating
the existence of long-ranged spin-triplet supercurrents in
ferromagnets proximity coupled to singlet superconductors open up
the possibility for new interesting physics and applications [for a
review, see M. Eschrig, Phys. Today \textbf{64}(1), 43 (2011)]. Our
group reported the injection of triplet supercurrents into a
magnetically uniform ferromagnet (Co) by sandwiching it between two
helimagnet/superconductor (Ho/Nb) bilayers to form a
Nb/Ho/Co/Ho/Nb-type Josephson device. In the function of the Ho
layer thicknesses, the supercurrent was found to modulate in a
complex way that seemed to depend on the magnetic structure of Ho.
To understand this unusual behavior, we have theoretically studied
the properties of an ideal Josephson device with a
helimagnet/ferromagnet/helimagnet (HM/F/HM) barrier in the clean
limit using the Eilenberger equation; we show, in particular, that
the maximum triplet supercurrent that can pass across the barrier
will depend non-monotonically on the thicknesses of the HM layers if
the HM and F layers are magnetically exchange coupled at their
interface.

\end{abstract}


\maketitle


\section{Introduction} \label{sec-int}

The intense interest in understanding the interplay between
superconductors (S) and ferromagnets (F) was primarily triggered by
the pioneering work of Ryazanov \emph{et al.},\cite{Ryazanov} who
showed that the supercurrent in a S/F/S junction containing the weak
ferromagnet CuNi was periodically modulated: first by temperature as
a result of varying the exchange energy of the low Curie temperature
CuNi, and also in the function of the CuNi thickness.\cite{Oboznov}
The oscillating and decay lengths of the thickness modulation are
much shorter ($\xi \approx$ 1 nm) if a strong ferromagnet is used,
such as Co,\cite{Co1Ni1,Co2,Co3} Fe,\cite{Fe1,Fe2} or
Ni,\cite{Co1Ni1,Ni2} but because of their long mean free paths for
both scattering and spin flip they are attractive materials to use
in S/F/S junctions. For a detailed review on the singlet proximity
coupling of S and F materials see Ref. \onlinecite{Buzdin} and
references therein.

The decay lengths of supercurrents in ferromagnets can be radically
extended if the electron pairs transform from a singlet to a triplet
spin state at a S/F interface via a spin-mixing process. This is
known as the long-range triplet proximity effect; for a
comprehensive review, see Ref. \onlinecite{BergeretRMP} and
references therein. One way to theoretically promote the spin-mixing
conversion between singlet pairs and triplet pairs at a S/F
interface is to incorporate a magnetically inhomogenous layer
between the S and F layers.\cite{BergeretPRL} The triplet electron
pairs which form are expected to be much more robust in
ferromagnets, with longer coherence lengths ($\xi \gg$ 1 nm) than
pairs in a singlet state. Therefore the demonstration of a
supercurrent in a S/F/S junction where the F layer thickness exceeds
any length scale possible for a singlet pair to exist is considered
indirect proof of the long-range triplet proximity effect.
Signatures of such a long-range proximity effect were first reported
more than two decades ago (see, e.g. Ref. \onlinecite{Lawrence}).
However, it was only in 2006 when the first major breakthrough was
achieved with supercurrents reported in planar Josephson devices
with half-metallic (i.e., fully spin polarized) barriers that were
hundreds of nanometers long;\cite{Keizer} see also the related
articles in Ref. \onlinecite{EschrigPRL}. In the same year, a
triplet superconducting state was also reported in Ho, a
helimagnetic rare-earth metal, which formed the junction of a
superconducting interferometer.\cite{Sosnin} These results
stimulated intense theoretical work aimed at understanding triplet
supercurrents better and how they could be created at S/F interfaces
in a more routine way.

Over the past year it would seem that the puzzle to control triplet
pair generation in S/F/S devices was finally solved with long-ranged
supercurrents consistent with spin-triplet theory demonstrated in a
wide range of ferromagnets,\cite{EschrigPT} including a Co/Ru/Co
synthetic antiferromagnet interfaced by normal (i.e., non-magnetic)
metal spacers and ferromagnetic alloys\cite{Khaire} (see also Ref.
\onlinecite{Trifunovic}); a Ho/Co/Ho composite
barrier\cite{Robinson} (see also Refs.
\onlinecite{Alidoust,Halasz,Usman}); a magnetic Cu$_2$MnAl Heusler
compound with a complex magnetic profile\cite{Sprungmann} (see also
Ref. \onlinecite{Linder}); a half-metallic CrO$_2$ wire;\cite{Anwar}
and a Co nanowire.\cite{Wang} In addition to these experiments,
superconducting gap features have been very recently measured by
scanning tunneling spectroscopy\cite{Kalcheim} in the half-metallic
manganite La$_{2/3}$Ca$_{1/3}$MnO$_3$ (LCMO) grown on
YBa$_2$Cu$_3$O$_{7-\delta}$. The gap features were observed in the
LCMO up to a thickness of approximately 30 nm, and so offer the
potential to explore the fundamental properties of a triplet
superconducting state.

The experiments reported in Refs. \onlinecite{Khaire},
\onlinecite{Robinson}, and \onlinecite{Sprungmann} share many
similarities with the theoretical triplet junction proposed by
Houzet and Buzdin:\cite{Houzet} a S/F$'$/F/F$''$/S junction in which
the magnetizations in the F$'$ and F$''$ layers are non-collinear
with the magnetization in the F layer, allowing control over the
creation of triplet electron pairs. Our group reported Josephson
devices in which the F$'$ and F$''$ layers were substituted by Ho
and the F layer used was Co.\cite{Robinson} The maximum supercurrent
(i.e., the Josephson critical current $I_C$) in these devices was
found to depend non-monotonically on the thickness $d_h$ of each Ho
layer, with peaks in $I_C$ when $d_h\approx$ 4.5 nm and $d_h\approx$
10 nm. These thicknesses appeared to correspond to an optimum
spin-triplet proximity effect, and by increasing the Co thickness
$d_f$ in the 0$-$16 nm range, a slow decay in $I_C$ was observed
with a decay length of $\sim$ 10 nm at 4.2 K. This decay length is
almost ten times larger than that found in simple (spin-singlet
dominated) Co-based Josephson devices.\cite{Co1Ni1,Co2} Alidoust and
Linder\cite{Alidoust} established theoretically that spin-triplet
supercurrents could account for the slow decay in $I_C$ with $d_f$,
but their model did not account for the complex dependence of $I_C$
on $d_h$. When rare-earth ferromagnets are grown on transition-metal
ferromagnets, it is well known that the two materials magnetically
couple at the interface. In view of this fact, we solve the
Eilenberger equation for an ideal S/HM/F/HM/S Josephson device with
arbitrary interfacial magnetic coupling at the HM/F interfaces, and
demonstrate a profound dependence of the triplet supercurrent
amplitude on the thicknesses of the HM layers. The results provide a
new level of understanding into the conversion process between
singlet and triplet Cooper pairing at a
superconductor/helimagnet/ferromagnet interface.

\section{General formalism} \label{sec-gen}

\begin{figure}[t]
\begin{center}
\includegraphics[width=7.0cm]{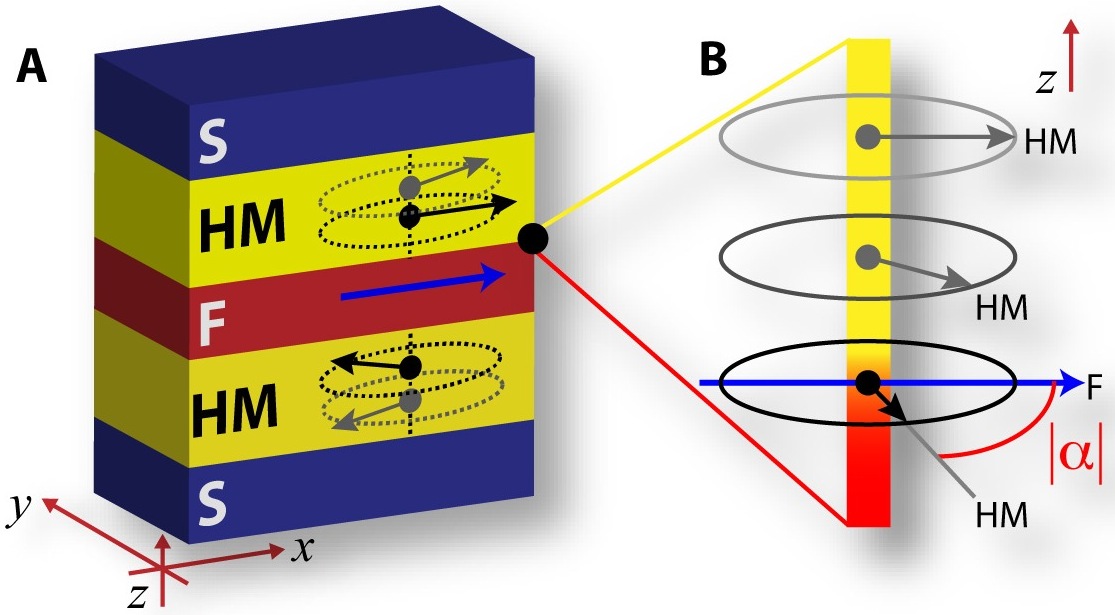}
\caption{(Color online) (A) Illustration of a superconducting
Josephson device with a helimagnet/ferromagnet/helimagnet barrier
(S/HM/F/HM/S). The HM layers control the conversion between singlet
and triplet Cooper pairing, while the F layer behaves like a spin
filter, only allowing triplet Cooper pairs to pass across it. The F
and HM layers are magnetically exchange coupled at the F/HM
interfaces (B) with an angle of $-\pi \leq \alpha \leq \pi$ radians
between the F and HM layer moments ($\alpha < 0$ in the figure).}
\label{fig-1}
\end{center}
\end{figure}

The S/HM/F/HM/S device considered in this paper is illustrated in
Fig. \ref{fig-1}. All layers are normal to the $z$ direction. The HM
layers have a thickness $d_h$ and their magnetization rotates in the
$\{ x,y \}$ plane as a function of $z$. \footnote{Here we neglect
the constant axial magnetization that is present in crystalline Ho
below 20K to simplify the mathematics; nevertheless, the same
conclusions are achieved if we include this component.} The strongly
ferromagnetic F layer has a thickness $d_f$ and a uniform
magnetization in the $+x$ direction. The S layers at $z < 0$ and $z
> d \equiv 2d_h + d_f$ are infinitely thick and in a singlet state.
Subscripts $h$, $f$, and $s$ refer to the HM, F, and S layers.

We assume a moderately clean limit and a temperature close to the
critical temperature $T_C$ of the superconducting leads, therefore
we adapt the linearized Eilenberger equation (LEE). Due to the
presence of non-uniform magnetization, we consider both singlet and
triplet correlations. The anomalous Green's function is then a $2
\times 2$ matrix given by $\hat{f} (z, \theta, \omega) = f^0 \,
\hat{1} + \vec{f} \cdot \vec{\hat{\sigma}}$, where $\vec{f} = (f^x,
f^y, f^z)$, and $\vec{\hat{\sigma}} = (\hat{\sigma}^x,
\hat{\sigma}^y, \hat{\sigma}^z)$ is a vector of the Pauli matrices.
The component $f^0$ is associated with spin-singlet pairs, while the
three components in $\vec{f}$ are associated with spin-triplet
pairs. To understand the meaning of the different triplet
components, we switch to an alternative representation where we have
$\hat{f} = f^0 \hat{1} + f^x \hat{\sigma}^x + f^{+} \hat{\sigma}^{+}
+ f^{-} \hat{\sigma}^{-}$ with $\hat{\sigma}^{\pm} = (\hat{\sigma}^y
\pm i\hat{\sigma}^z) / \sqrt{2}$ and $f^{\pm} = (f^y \mp i f^z) /
\sqrt{2}$. The components $f^x$ and $f^{\pm}$ then correspond to
triplet pairs with spin projections $0$ and $\pm 1$ to the $x$
direction, respectively.

Since the HM layers have a weak magnetization, all components in
$\hat{f}$ are non-zero. The Fermi surface average is $\langle
\hat{f} \rangle \approx 0$ in the moderately clean
limit\cite{Konschelle} and so the LEE reads
\begin{equation}
v_h \cos \theta \, \frac{\partial \hat{f}_h} {\partial z} + \left(
2\omega + \tau_h^{-1} \right) \hat{f}_h + i \vec{I}_h \cdot \{
\vec{\hat{\sigma}}, \hat{f}_h \} = 0, \label{eq-gen-hm}
\end{equation}
\noindent where $\{ x,y \} \equiv xy + yx$ is the anticommutator,
while $v_a$ and $\tau_a$ are the Fermi velocity and the
pair-breaking time in a generic layer $a$. The electron mean free
path is then $\ell_a = v_a \tau_a$. The Matsubara frequencies are
given by $\omega = \pi k_B T (1 + 2n) / \hbar$, and $\vec{I}_h$ is
the magnetic exchange field in units of frequency. This field takes
the form $\vec{I}_h = I_h (\cos [Q \tilde{z} + \alpha], \sin [Q
\tilde{z} + \alpha], 0)$ in the HM layers, where $\tilde{z} \equiv
|z - d/2| - d_f/2$ is the distance to the nearest HM/F interface and
$Q$ is the wave vector of the magnetic helix. The magnetic exchange
coupling at the HM/F interfaces is included by introducing an
anisotropy angle $\alpha$ between the local magnetizations of the HM
and F layers. When $\alpha$ is 0 or $\pi$, the magnetizations are
locally parallel or antiparallel, respectively. \footnote{Note that
when $\alpha < 0$ the HM magnetization rotates too much as
$\tilde{z} \rightarrow 0$ and when $\alpha > 0$ it does not rotate
enough. The symmetry in the definition of $\tilde{z}$ assumes that
the helicities of the two HM layers are opposite. It turns out that
reversing the helicity of one layer only gives an overall minus sign
in the expression for the critical current, which is not
measurable.}

For the strongly magnetized F layer there are two independent spin
bands for the majority ($+x$) and the minority ($-x$) spins, and the
singlet correlations are destroyed on a sub-nanometer scale.
Therefore we take $f^0 = f^x = 0$ and identify the remaining
components $f^{\pm}$ with triplet pairs from the majority and
minority spin bands. The governing equations for the two independent
spin bands are then
\begin{equation}
v_f^{\pm} \cos \theta \, \frac{\partial f_f^{\pm}} {\partial z} +
\left( 2\omega + \tau_f^{-1} \right) f_f^{\pm} = 0,
\label{eq-gen-fm}
\end{equation}
\noindent where we assume $v_f^{+} > v_f^{-}$. However, since
scattering is mainly due to impurities at low temperatures, we
assume that the pair-breaking time $\tau_f$ is the same for both
spin bands.

At the interface $z = z_{ab}$ between two generic layers $a$ and
$b$, the parallel component of the Fermi momentum is conserved. If
we assume that all bands are parabolic with the same effective
electron mass, the Fermi momentum is proportional to the Fermi
velocity, meaning $\hat{f}_a (z_{ab}, \theta_a, \omega) = \hat{f}_b
(z_{ab}, \theta_b, \omega)$ for $v_a \sin \theta_a = v_b \sin
\theta_b$. If we further assume that $v_s \ll v_h,v_f$, it is valid
to use the single-channel approximation; the anomalous Green's
function $\hat{f}$ is then only non-zero for angles close to $0$ or
$\pi$ in the HM and F layers, therefore $\cos \theta$ can be
substituted by $\pm1$ in Eqs. (\ref{eq-gen-hm}) and
(\ref{eq-gen-fm}).

The phase difference between the two S layers is $\Phi$ and the
magnitude $\Delta$ of the bulk pairing potentials ($\Delta$ and
$\Delta e^{i \Phi}$) is determined from the self-consistency
equation. For $T \approx T_C$, we can apply rigid boundary
conditions at the S/HM interfaces.\cite{Houzet} This means that
$\hat{f}$ is the same at the interface as in the bulk of the S
layers for outgoing directions. The boundary conditions at the S/HM
interfaces are then $\hat{f}_s (0, \theta, \omega) = F(\Delta,
\omega) \, \hat{1}$ for $\cos \theta
> 0$ and $\hat{f}_s (d, \theta, \omega) = F(\Delta, \omega) \, e^{i
\Phi} \,\hat{1}$ for $\cos \theta < 0$, where $F(\Delta, \omega) =
\Delta / \sqrt{\Delta^2 + \hbar^2 \omega^2}$ is the equilibrium
value of $f^0$ and the subscript $s$ indicates that the expressions
correspond to the S side of the interface.

In the formalism of the LEE, the Josephson current density at a
generic position $z$ in the junction is given by
\begin{equation}
J = \frac{3\pi k_B T \sigma_z} {2e \ell_z} \sum_{\omega > 0} \int
\frac{d\Omega} {4\pi} \cos \theta \, \mathrm{Im} \, \mathrm{Tr}
\left[ \hat{f}^{\dag} (z, \theta, \omega) \hat{f}' (z, \theta',
\omega) \right], \label{eq-gen-curr}
\end{equation}
\noindent where $\theta' \equiv \pi - \theta$, and $\sigma_z$ is the
normal-state conductivity of the layer at position $z$. The function
$\hat{f}'$ satisfies the same equations as $\hat{f}$ with the same
boundary conditions at the interfaces, reversed exchange field
($-\vec{I}_h$) in the HM layers, and exchanged spin bands ($v_f^{+}
\leftrightarrow v_f^{-}$) in the F layer.

\section{Critical current of the junction} \label{sec-curr}

In this section we calculate the Josephson current at the S side of
the $z = d$ interface and hence obtain the critical current of the
junction. Since $\hat{f}_s (d, \theta, \omega)$ for $\cos \theta <
0$ is known and only its $f^0$ component is non-zero, the Josephson
current is determined entirely by $f_s^0 (d, \theta, \omega)$ for
$\cos \theta > 0$. Due to the linearity of Eqs. (\ref{eq-gen-hm})
and (\ref{eq-gen-fm}), we can write
\begin{equation}
f_s^0 (d, \theta, \omega) = S(\theta, \omega) f_s^0 (0, \theta,
\omega) = S(\theta, \omega) F(\Delta, \omega), \label{eq-curr-S-1}
\end{equation}
\begin{equation}
S(\theta,\omega) = \mathbf{S}_h^T \cdot \mathbf{S}_f \cdot
\mathbf{S}_h, \label{eq-curr-S-2}
\end{equation}
\noindent where the different matrices/vectors $\mathbf{S}$ describe
the effects of the different layers between $z = 0$ and $z = d$.

The role played by the HM layers is to convert between the singlet
component $f^0$ in the S layers and the triplet components $f^{\pm}$
in the F layer, implying $\mathbf{S}_h$ is a $2 \times 1$ vector,
while $\mathbf{S}_h^T$ is a $1 \times 2$ vector. The expressions for
$\mathbf{S}_h$ and $\mathbf{S}_h^T$ can be obtained by solving Eq.
(\ref{eq-gen-hm}) in the two HM layers. The vector corresponding to
the lower HM is
\begin{equation}
\mathbf{S}_h = -i B \exp (-D_h) \left( \begin{array}{c} 1 \\ 1
\end{array} \right), \label{eq-curr-S-h}
\end{equation}
\noindent
\begin{equation}
B = \frac{q \cos \alpha (1 - \cos \delta_h) + \sqrt{1 + q^2} \sin
\alpha \sin \delta_h} {\sqrt{2} (1 + q^2)}, \label{eq-curr-B}
\end{equation}
\noindent where $q \equiv Q \xi_h$, and $\xi_h = v_h / 2I_h$ is the
correlation oscillating length in the HM layers. We also define
reduced thicknesses as $\delta_h \equiv d_h \sqrt{1 + q^2} / \xi_h$
and $D_a \equiv d_a (2\omega + \tau_a^{-1}) / v_a$, where the latter
one is valid for a generic layer $a$. Physically, the parameter $B$
describes the conversion efficiency between singlet and triplet
Cooper pairs in each HM layer, while $B^2$ describes the total
efficiency of both HM layers.

Since the non-zero triplet components $f^{\pm}$ are independent in
the F layer, $\mathbf{S}_f$ is a $2 \times 2$ diagonal matrix.
Integrating Eq. (\ref{eq-gen-fm}) gives
\begin{equation}
\mathbf{S}_f = \left( \begin{array}{cc} \exp(-D_f^{+}) & 0 \\ 0 &
\exp(-D_f^{-}) \end{array} \right). \label{eq-curr-S-f}
\end{equation}
\noindent The fact that $v_f^{-} < v_f^{+}$ implies $D_f^{-} >
D_f^{+}$, therefore the triplet component $f^{-}$ corresponding to
the minority spin band decays faster than $f^{+}$.

Equations (\ref{eq-curr-S-h})$-$(\ref{eq-curr-S-f}) were derived by
setting $\cos \theta \approx 1$ in the HM and F layers.
Consequently, $S(\omega) \equiv S(\theta, \omega)$ is independent of
$\theta$. If we substitute Eq. (\ref{eq-curr-S-1}) and $\hat{f}_s
(d, \theta, \omega) = F(\Delta, \omega) \, e^{i \Phi} \,\hat{1}$
(for $\cos \theta < 0$) into Eq. (\ref{eq-gen-curr}), we recover the
usual current-phase relation, $J = J_C \sin \Phi$, and the critical
current density becomes
\begin{equation}
J_C = \frac{3\pi k_B T \sigma_s} {e \ell_s} \int_0^1 d\zeta \cdot
\zeta \, \sum_{\omega > 0} F(\Delta, \omega)^2 \, \mathrm{Re} \left[
S(\omega) \right]. \label{eq-curr-JC}
\end{equation}
The integral in $\zeta = \cos \theta$ gives $1/2$, while the sum in
$\omega$ requires a further approximation. Since $\omega \ll
\tau_a^{-1}$ for all layers and all Matsubara frequencies with a
significant contribution, we can neglect $\omega$ next to
$\tau_a^{-1}$ in $D_a$. This implies that $D_a \approx d_a / \ell_a$
and that $S \equiv S(\omega)$ is independent of $\omega$ as well.
The sum is now evaluated using $\sum_{n = 0}^{\infty} \frac{1} {p^2
+ (1 + 2n)^2} = \frac{\pi}{4p}\tanh \left( \frac{p \pi}{2} \right)$
and so the characteristic voltage reads
\begin{equation}
I_C R_N = \frac{3\pi \Delta \sigma_s} {8 e \ell_s} \tanh \left(
\frac{\Delta} {2 k_B T} \right) \left( \frac{2d_h} {\sigma_h} +
\frac{d_f} {\sigma_f} \right) \, S, \label{eq-curr-IC-1}
\end{equation}
\begin{equation}
S = -B^2 \exp \left( - \frac{2d_h} {\ell_h} \right) \sum_{\pm} \exp
\left( - \frac{d_f} {\ell_f^{\pm}} \right). \label{eq-curr-IC-2}
\end{equation}
The two terms in the sum are contributions from triplet pairs
traveling through the majority and minority spin bands of the F
layer, respectively.

\section{Discussion} \label{sec-disc}

Since only triplet pairs can enter the strongly ferromagnetic F
layer, the conversion between singlet and triplet pairs in the HM
layers is crucial. This process is represented by the efficiency
parameter $B^2$ in Eq. (\ref{eq-curr-IC-2}) and in this section we
discuss how the conversion depends on $d_h$, $q$, and $\alpha$.

\begin{figure}[t]
\begin{center}
\includegraphics[width=7.0cm]{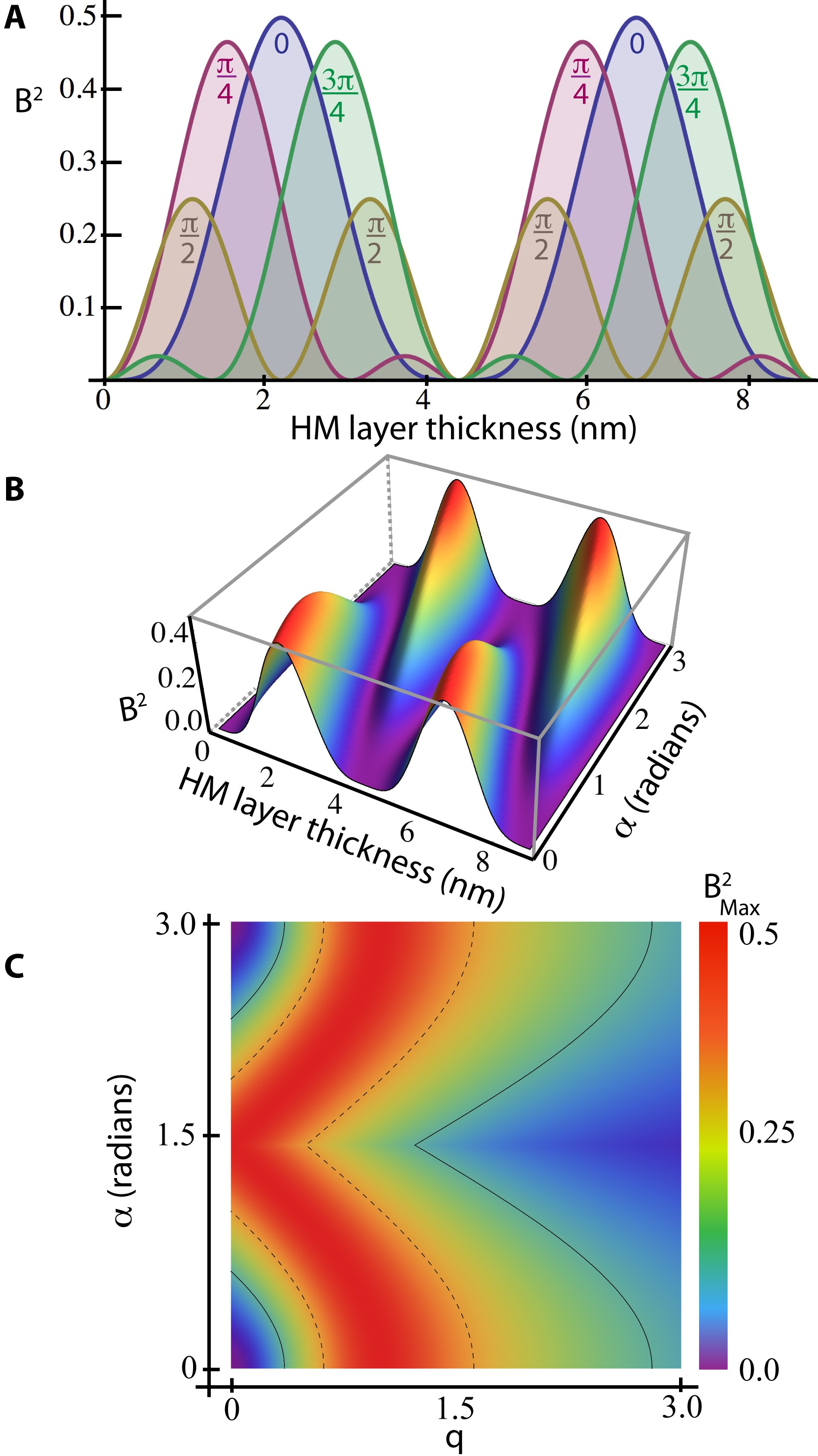}
\caption{(Color online) The efficiency parameter $B^2$ peaks at
particular HM layer thicknesses (A), and the positions and
magnitudes of these peaks depend on the magnetic anisotropy angle
$\alpha$ between neighboring HM and F layer moments (A, B). In (A)
and (B), $q = 1$. For optimum HM layer thicknesses, the maximum
efficiency parameter $B^2_{\mathrm{max}}$ depends sensitively on $q$
and $\alpha$ (C).} \label{fig-2}
\end{center}
\end{figure}

We first note from Eq. (\ref{eq-curr-B}) that $B^2$ is $\pi$
periodic in the anisotropy angle $\alpha$. This implies that the
critical current does not distinguish between locally parallel and
antiparallel magnetizations of the HM and F layers. In both of these
cases, the efficiency parameter takes the form
\begin{equation}
B^2 = \frac{2q^2} {(1 + q^2)^2} \sin^4 \left( \frac{d_h \sqrt{1 +
q^2}} {2\xi_h} \right) \quad (\alpha = 0, \pi), \label{eq-disc-B-1}
\end{equation}
which shows that the periodicity of $B^2$ in $d_h$ is in general
$\Delta d_h = 2\pi \xi_h / \sqrt{1 + q^2}$. The periodicity $\Delta
d_h$ depends on the two intrinsic length scales of the HM layers:
the oscillating length $\xi_h$ and the wavelength $\lambda = 2\pi /
Q$ of the magnetic helix. Furthermore, Eq. (\ref{eq-disc-B-1}) shows
that the shorter length scale is the dominant one. When the
oscillating length is small (i.e., $q \ll 1$), we have $\Delta d_h
\approx 2\pi \xi_h$ and so the helical structure becomes irrelevant.
The periodicity is determined by $\xi_h$ as for uniform magnets.
When the oscillating length is large (i.e., $q \gg 1$), we have
$\Delta d_h \approx \lambda$ and so that the periodicity is
determined entirely by the helical structure. It also follows from
Eq. (\ref{eq-disc-B-1}) that the zeros of $B^2$ are at $d_h = m \,
\Delta d_h$, and that its maxima are halfway between its zeros.

If we start increasing $\alpha$, the periodicity $\Delta d_h$
remains unchanged, and $B^2$ still has zeros at $d_h = m \, \Delta
d_h$. However, the primary maxima halfway are shifted to the left,
and secondary maxima appear with additional zeros in between (see
Fig. \ref{fig-2}). When the local HM and F magnetizations are
orthogonal, the efficiency parameter reads
\begin{equation}
B^2 = \frac{1} {2(1 + q^2)} \sin^2 \left( \frac{d_h \sqrt{1 + q^2}}
{\xi_h} \right) \quad (\alpha = \pm \pi/2).  \label{eq-disc-B-2}
\end{equation}
The primary and secondary maxima become equivalent, the periodicity
in $d_h$ is reduced by $2$, and the zeros of $B^2$ occur at $d_h = m
\, \Delta d_h / 2$.

It is interesting to look at the efficiency of the singlet/triplet
conversion at optimal HM layer thicknesses. This is determined by
$q$ and $\alpha$. In the limit when $q \gg 1$, the critical current
vanishes. This is because if $\lambda$ is too small, the HM
magnetization averages to zero within the length scale of $\xi_h$.
In the opposite limit when $q \ll 1$, we need $\alpha \neq 0$ for
$I_C \neq 0$. If $\lambda$ is too large, the HM layers become
uniformly magnetized, and we recover the results in Ref.
\onlinecite{Houzet} as $I_C \propto B^2 \propto \sin^2 \alpha$. To
be more quantitative, we maximize $B^2$ with respect to $d_h$; this
calculation gives
\begin{equation}
B_{\mathrm{max}}^2 = \frac{\left( q |\cos \alpha| + \sqrt{q^2 +
\sin^2 \alpha} \right)^2} {2(1 + q^2)^2}, \label{eq-disc-B-max}
\end{equation}
and the dependence of $B_{\mathrm{max}}^2$ on $q$ and $\alpha$ is
shown in Fig. \ref{fig-2}(C). We can establish that the most
efficient singlet/triplet conversion with $B_{\mathrm{max}}^2 = 1/2$
occurs whenever $q = |\cos \alpha|$. This is in fact an absolute
theoretical maximum. Due to the two triplet channels $f^{\pm}$ in
the F layer, $B_{\mathrm{max}}^2 = 1/2$ corresponds to perfect
conversion.

\begin{figure}[t]
\begin{center}
\includegraphics[width=7.5cm]{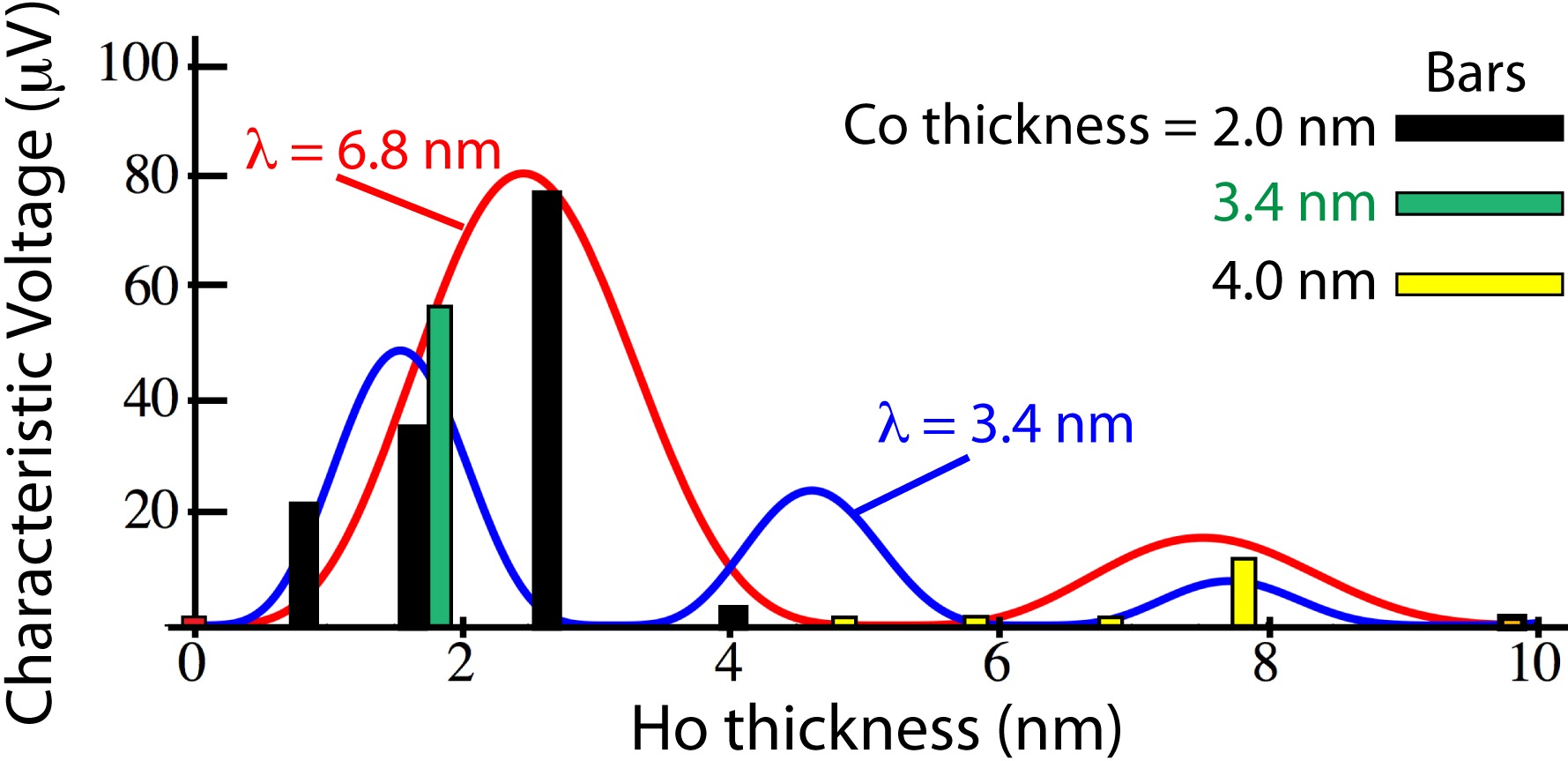}
\caption{(Color online) Experimental (bars, adapted from Ref.
\onlinecite{Robinson}) and theoretical (curves) $I_C R_N$ values of
a Nb/Ho/Co/Ho/Nb device against the symmetrical thickness of Ho at
$4.2$ K. Theoretical curves are plotted for $\lambda =$ 3.4 nm
(blue) and $\lambda$ = 6.8 nm (red), while the Co thickness is kept
at 3.4 nm (note that the local magnitude of $I_C R_N$ only weakly
depends on the Co thickness). For the experimental $I_C R_N$ values,
the magnetically dead layers present at each Ho surface ($\approx$
1.2 nm) have been subtracted from the total Ho thickness.
Theoretical curves are plotted with the following realistic
parameter values: $T_C =$ 9.1 K, $v_s$ = 0.4 $\times$ 10$^6$
ms$^{-1}$, $v_h = v_f^{+} =$ 1.0 $\times$ 10$^6$ ms$^{-1}$, $v_f^{-}
=$ 0.75 $\times$ 10$^6$ ms$^{-1}$, $\ell_h =$ 4 nm, $\ell_f^{+} =$
12 nm, $\ell_f^{-} =$ 9 nm, $I_h =$ 0.25 eV, and $\alpha = 0$.}
\label{fig-3}
\end{center}
\end{figure}

Altogether, the dependence of $B^2$ on the HM layer thickness shown
in this paper demonstrates the profound effect that interfacial
layer-by-layer magnetic coupling can have on the amplitude of the
triplet supercurrent in a S/HM/F/HM/S device. Even when the
magnetizations at the HM/F interfaces couple parallel or
antiparallel, the amplitude of the triplet supercurrent oscillates
with peaks and zeros commensurate on $d_h$. This result agrees with
experimental $I_C$ data reported in Ref. \onlinecite{Robinson}: in
Fig. \ref{fig-3} we directly compare the theoretical and
experimental dependence of $I_C R_N$ on Ho layer thickness and
agreement is achieved for $\lambda$ in the 3.4$-$6.8 nm range,
consistent with $\lambda$ values estimated in Ho thin
films.\cite{Leiner}

This close agreement between theory and experiment provides a
clearer understanding of the role played by interfacial magnetic
coupling between Ho and Co on the $I_C$ found in Nb/Ho/Co/Ho/Nb
devices.\cite{Robinson} The possibility of controlling $I_C$ by
manipulating the coupling anisotropy (i.e., $\alpha$) in these or
similar devices has not been explored experimentally so far. Since
the Curie temperature of Co ($\sim$ 1000 K) is significantly higher
than either the N\'{e}el ($\sim$ 130 K) or the Curie ($\sim$ 20 K)
temperature of Ho, it may be possible to vary $\alpha$ by field
cooling a device from $\sim$ 130 K. By repeating this procedure with
the field applied at various in-plane angles, the effect of $\alpha$
on the amplitude of the triplet supercurrent $I_C$ could be tested
with potentially large $\Delta I_C/I_C$ ratios obtainable.

\section{Summary} \label{sec-sum}

Layer-by-layer interfacial magnetic coupling between rare-earth
helimagnets (HM) and ferromagnets (F) can strongly influence the
interconversion between singlet and triplet Cooper pairing in
S/HM/F/HM/S-type Josephson devices. This results in the amplitude of
the spin-triplet supercurrent passing through the F layer depending
sensitively on the magnetic structure and the thicknesses of the
helimagnetic layers.


\begin{acknowledgments}

We are grateful to F. S. Bergeret, A. I. Buzdin, F. Chiodi, and J.
Linder for valuable advice during the preparation of this paper. The
research was funded by St. John's College, Cambridge, and the UK
EPSRC.

\end{acknowledgments}


\end{document}